\documentclass[english,aps,preprint]{revtex4}
\usepackage[T1]{fontenc}
\usepackage[latin9]{inputenc}
\usepackage{float}
\usepackage{amssymb}
\usepackage{graphicx}

\makeatletter

\providecommand{\tabularnewline}{\\}

\@ifundefined{textcolor}{}
{%
 \definecolor{BLACK}{gray}{0}
 \definecolor{WHITE}{gray}{1}
 \definecolor{RED}{rgb}{1,0,0}
 \definecolor{GREEN}{rgb}{0,1,0}
 \definecolor{BLUE}{rgb}{0,0,1}
 \definecolor{CYAN}{cmyk}{1,0,0,0}
 \definecolor{MAGENTA}{cmyk}{0,1,0,0}
 \definecolor{YELLOW}{cmyk}{0,0,1,0}
}

\makeatother

\usepackage{babel}
\begin{document}

\title{Search for excited spin-3/2 neutrinos at LHeC}

\author{A. Ozansoy}

\email{aozansoy@science.ankara.edu.tr}

\selectlanguage{english}%

\affiliation{Ankara University, Department of Physics, 06100 Tandogan, Ankara,
Turkey}

\author{V. Ar\i{}}

\email{vari@science.ankara.edu.tr}

\selectlanguage{english}%

\affiliation{Ankara University, Department of Physics, 06100 Tandogan, Ankara,
Turkey}

\author{V. \c{C}etinkaya}

\email{volkan.cetinkaya@dpu.edu.tr}

\selectlanguage{english}%

\affiliation{Dumlupinar University, Department of Physics, 43100 Merkez, Kutahya,
Turkey}

\begin{abstract}
We study the potential of the next $ep$ collider, namely LHeC, with
two options $\sqrt{s}=1.3$ TeV $\sqrt{s}=1.98$ TeV, to search for
excited spin-1/2 and spin-3/2 neutrinos. We calculate the single production
cross section of excited spin-1/2 and spin-3/2 neutrinos according
to their effective currents describing their interactions between
gauge bosons and SM leptons. We choose the $\nu^{\star}\rightarrow eW$
decay mode of excited neutrinos and $W\rightarrow jj$ decay mode
of $W$-boson for the analysis. We put some kinematical cuts for the
final state detectable particles and plot the invariant mass distributions
for signal and the corresponding backgrounds. In order to obtain accessible
limits for excited neutrino couplings, we show the $f-f^{\prime}$
and $c_{iV}-c_{iA}$ contour plots for excited spin-1/2 and excited
spin-3/2 neutrinos, respectively.
\end{abstract}
\maketitle

\section{introduction}

The Standard Model (SM) of the particle physics is in accordance with
the experimental outcomes received from the operating colliders. The
first run of the Large Hadron Collider (LHC) brought the expected
Higgs boson discovery, so a crucial part of the SM had been completed.
But there is still no satisfying answer about the three-family structure
of leptons and quarks and mass hierarchy of them. An attractive explanation
is lepton and quark compositeness \cite{mycitation-1}. In composite
models, known leptons and quarks have a substructure characterized
by an energy scale named compositeness scale, $\Lambda.$ A natural
consequence of compositeness is the occurrence of excited states \cite{mycitation-2,mycitation-3}.
Phenomenologicaly, an excited lepton can be regarded as a heavy lepton
sharing the same leptonic quantum number with the corresponding SM
lepton. If leptons present composite structures, they can be considered
as spin-1/2 bound states containing three spin-1/2 or spin-1/2 and
spin-0 subparticles. Bound states of spin-3/2 leptons also possible
with three spin-1/2 \cite{mycitation-1} or spin-1/2 and spin-1 subparticles
in the framework of composite models \cite{mycitation-4}. The motivations
for spin-3/2 particles come from two different scenarios; spin-3/2
leptons appear in composite models \cite{mycitation-5,mycitation-6};
and a spin-3/2 gravitino is the superpartner of graviton in supergravity\cite{mycitation-7}.
One can find some of the latest studies about beyond the Standard
Model theories including exotic spin-3/2 particles in \cite{mycitation-8-1}.

Both excited spin-1/2 and spin-3/2 neutrinos can be produced at future
high energy lepton, hadron and lepton-hadron colliders. Elaborate
studies on excited spin-1/2 neutrinos can be found in \cite{mycitation-9,mycitation-10,mycitation-11,mycitation-12,mycitation-8}.
Also, one can find excited spin-1/2 neutrino production by ultra high
energy neutrinos in \cite{mycitation-13} and the impact of excited
spin-1/2 neutrinos on $\nu\bar{\nu}\rightarrow\gamma\gamma$ process
in \cite{mycitation-14}.

The mass limit for excited spin-1/2 neutrinos obtained from their
pair production ($e^{+}e^{-}\rightarrow\nu^{\star}\nu^{\star}$ process)
by L3 Collaboration at $\sqrt{s}=189-209$ GeV, assuming $f=-f^{\prime}$,
where $f$ and $f^{\prime}$ are the new couplings determined by the
composite dynamics, is $m^{\star}>102.6$ GeV \cite{mycitation-15}.
Assuming $f=f^{\prime}$ and $f/\Lambda=1/m^{\star}$, for single
production of excited spin-1/2 neutrino in $ep$ collisions taking
into account all the decay channels, the H1 Collaboration set the
exclusion limit for the mass range of excited neutrino $m^{\star}>213$
GeV at 95\% C.L.\cite{mycitation-16}. Recently, a search was performed
by the ATLAS Collaboration taking into account pair production of
excited spin-1/2 neutrinos either through contact or gauge-mediated
interactions and their decay proceeds via the same mechanism. Considering
events with at least three charged leptons with $\Lambda=m^{\star}$,
$f=f^{\prime}=1$ and with an integrated luminosity of $20.3$ $fb^{-1}$
of $pp$ collisions at $\sqrt{s}=8$ TeV; lower mass limit obtained
as 1.6 TeV for every excited spin-1/2 neutrino flavour\cite{mycitaton-17}.

Excited spin-3/2 neutrinos were least studied in the litterateur by
the side of the spin-1/2 ones. An investigation for the production
and decay process of the single heavy spin-3/2 neutrino was performed
in \cite{mycitation-18,my citation-19}. The study for the potential
of future high energy $e^{+}e^{-}$ linear colliders to probe excited
spin-3/2 neutrino signals in different decay modes in the frame of
three phenomenological currents taking into account the corresponding
background was done in \cite{mycitation-4}.

Studies are ongoing for the development of a new $ep$ collider, the
Large Hadron Electron Collider (LHeC), with an electron beam of 60
GeV, to possibly 140 GeV, and a proton beam of the LHC \cite{mycitation-20}
or in the future the Future Circular Collider lepton-hadron collider
(FCC-eh)\cite{mycitation-21}. The LHeC is the highest energy lepton-hadron
collider under design and is considered as a linac-ring collider.
Linac-ring type colliders were proposed in \cite{mycitation-22} and,
the physics potentials and advantages of these type lepton-hadron
colliders are discussed in \cite{mycitaiton-23}. Latest results for
excited neutrino searches coming from the first $ep$ collider HERA
have showed that $ep$ colliders are so competitive to $pp$ and $e^{+}e^{-}$
colliders and very important for the investigation of beyond SM physics
\cite{mycitation-16,mycitation-20}. With the design luminosity of
$10^{33\:}cm^{-2}s^{-1}$ the LHeC is intended to exceed the HERA
luminosity by a factor of $\sim100$. So it would be a major opportunity
to push forward the investigations done in the LHC.

This work is a continuation of the previous works on excited neutrinos
\cite{mycitation-4,mycitation-10}. In this work, in Section II we
introduce the phenomenological currents for excited neutrinos and
give the decay widths of them. In Section III, we consider single
production of excited spin-1/2 and spin-3/2 neutrinos at $ep$ colliders.
We take into account the signal in $\nu^{\star}\rightarrow eW$ decay
mode of excited neutrinos as well as corresponding backgrounds at
LHeC with $\sqrt{s}=1.3$ TeV and $\sqrt{s}=1.98$ TeV. We plot the
invariant mass distributions for single production of excited neutrinos
with spin-1/2 and spin-3/2. Last, we plot the contour plots for the
excited neutrino couplings to obtain the exclusion limits. Investigestion
on excited fermions with spin-1/2 take an important part in the physics
program of LHeC. Although the latest limit for excited spin-1/2 neutrinos
set by the ATLAS experiment is high, it is important to examine the
excited neutrinos with different spins at high energy lepton-hadron
colliders. This work is the only dedicated work which gives the comparative
results both for excited spin-1/2 and spin-3/2 neutrinos to comprehend
the potential of next $ep$ collider.

\section{Physical Preliminaries}

An excited spin-1/2 neutrino is the lowest radial and orbital excitation
according to the classification by $SU(2)\times U(1)$ quantum numbers.
Interactions between excited spin-1/2 neutrino and ordinary leptons
are magnetic transition type \cite{mycitation-24,mycitation-25,mycitation-26}.
The effective current for the interaction between an excited spin-1/2
neutrino, a gauge boson ($V=\gamma,Z,W^{\pm}$), and the SM lepton
is given by

\begin{equation}
J^{\mu}(1/2)=\frac{g_{e}}{2\Lambda}\overline{u}(k,1/2)i\sigma^{\mu\nu}q\nu(1-\gamma_{5})f_{V}u(p,1/2)
\end{equation}

\noindent where $\Lambda$ is the new physics scale; $g_{e}$ is electromagnetic
coupling constant with $g_{e}=\sqrt{4\pi\alpha}$; $k,p$ and $q$
are the four momentum of the SM lepton, excited spin-1/2 neutrino
and the gauge boson, respectively. $f_{V}$ is the new electroweak
coupling parameter corresponding to the gauge boson $V$ and $\sigma^{\mu\nu}=i(\gamma^{\mu}\gamma^{\nu}-\gamma^{\nu}\gamma^{\mu})/2$
with $\gamma^{\mu}$ being the Dirac matrices. An excited neutrino
has three possible decay modes each of one is related to a vector
boson $\gamma,W$ and $Z$; radiative decay $\nu^{\star}\rightarrow\nu\gamma$,
neutral weak decay $\nu^{\star}\rightarrow\nu Z$ , and charged weak
decay $\nu^{\star}\rightarrow eW$. Neglecting SM lepton mass we find
the decay width of excited spin-1/2 neutrino as

\begin{equation}
\Gamma(l^{\star(1/2)}\rightarrow lV)=\frac{\alpha m^{\star3}}{4\Lambda^{2}}f_{V}^{2}(1-\frac{m_{V}^{2}}{m^{\star2}})^{2}(1+\frac{m_{V}^{2}}{2m^{\star2}})
\end{equation}

\noindent where $f_{\gamma}=(f-f^{\prime})/2,$$f_{Z}=(fcot\theta_{W}+f^{\prime}tan\theta_{W})/2,$$f_{W}=f/\sqrt{2}sin\theta_{W}$;
$\theta_{W}$ is the weak mixing angle and $m_{V}$ is the mass of
the gauge boson. The couplings $f$ and $f^{\prime}$ are the scaling
factors for the gauge couplings of $SU(2)$and $U(1)$. It is remarkable
that for the choice of the couplings $f=-f^{\prime}$, the electromagnetic
interaction of excited neutrino and SM neutrino exists. Brancing ratios
of excited spin-1/2 neutrino are presented in Table I. One may note
that for the choice $f=-f^{\prime}=1$ the brancing ratio for the
$eW$ channel is $~60$\%. Hence, to choose the $\nu^{\star}\rightarrow eW$
mode for the analysis is more feasible.

\begin{table}[H]
\caption{Brancing ratios and total decay width of excited spin-1/2 neutrinos
for $f=-f^{\prime}=1\:(f=f^{\prime}=1)$ Here it is taken $\Lambda=m^{\star}$.}

\begin{tabular}{|c|c|c|c|c|}
\hline 
$m^{\star}(GeV)$ & $\Gamma(GeV)$ & $\%BR($$\nu^{\star}\rightarrow\nu\gamma)$ & $\%BR($$\nu^{\star}\rightarrow\nu Z$$)$ & $\%BR($$\nu^{\star}\rightarrow eW$)\tabularnewline
\hline 
\hline 
300 & 1.91 & 30.5 (0) & 10.7 (38.3) & 58.9 (61.7)\tabularnewline
\hline 
500 & 3.36 & 28.9 (0) & 11.1 (38.9) & 60.0 (61.1)\tabularnewline
\hline 
750 & 5.12 & 28.4 (0) & 11.3 (39.0) & 60.3 (61.0)\tabularnewline
\hline 
1000 & 6.87 & 28.2 (0) & 11.3 (39.1) & 60.4 (60.9)\tabularnewline
\hline 
1500 & 10.35 & 28.1 (0) & 11.4 (39.1) & 60.5 (60.9)\tabularnewline
\hline 
2000 & 13.82 & 28.1 (0) & 11.4 (39.1) & 60.5 (60.9)\tabularnewline
\hline 
2500 & 17.28 & 28.1 (0) & 11.4 (39.1) & 60.5 (60.9)\tabularnewline
\hline 
3000 & 20.75 & 28.1 (0) & 11.4 (39.1) & 60.5 (60.9)\tabularnewline
\hline 
\end{tabular}
\end{table}

The two phenomenological currents for the interactions between an
excited spin-3/2 neutrino, a gauge boson ($V=\gamma,Z,W^{\pm}$),
and the SM lepton are given by

\begin{equation}
J_{1}^{\mu}(3/2)=g_{e}\overline{u}(k,1/2)(c_{1V}-c_{1A}\gamma_{5})u^{\mu}(p,3/2),
\end{equation}

\begin{equation}
J_{2}^{\mu}(3/2)=\frac{g_{e}}{\Lambda}\overline{u}(k,1/2)q_{\lambda}\gamma^{\mu}(c_{2V}-c_{2A}\gamma_{5})u^{\lambda}(p,3/2),
\end{equation}

Decay widths of excited spin-3/2 neutrinos for the $\nu^{\star}\rightarrow\nu\gamma$
decay mode for the two currents are given by

\begin{equation}
\Gamma_{1}(\nu^{\star(3/2)}\rightarrow\nu\gamma)=\frac{\alpha}{4}(c_{1v}^{\gamma^{2}}+c_{1A}^{\gamma^{2}})m^{\star},
\end{equation}

\begin{equation}
\Gamma_{2}(\nu^{\star(3/2)}\rightarrow\nu\gamma)=\frac{\alpha}{24}(c_{2v}^{\gamma^{2}}+c_{2A}^{\gamma^{2}})m^{\star}(\frac{m^{\star}}{\Lambda})^{2},
\end{equation}

\noindent and for the neutral and charged weak decay modes ($\nu^{\star}\rightarrow\nu Z$
and $\nu^{\star}\rightarrow eW$) given as

\begin{equation}
\Gamma_{1}(\nu^{\star(3/2)}\rightarrow lV)=\frac{\alpha}{48}(c_{1v}^{2}+c_{1A}^{2})m^{\star}\frac{(1-\kappa)^{2}}{\kappa}(1+10\kappa+\kappa^{2}),
\end{equation}

\begin{equation}
\Gamma_{2}(\nu^{\star(3/2)}\rightarrow lV)=\frac{\alpha}{48}(c_{2v}^{2}+c_{2A}^{2})m^{\star}(\frac{m^{\star}}{\Lambda})^{2}\frac{(1-\kappa)^{4}}{\kappa}(1+2\kappa),
\end{equation}

\noindent where $\kappa=(m_{V}/m_{\star})^{2}$, $V=Z,W,$ and $l=e,\nu$. Branching ratios and total decay width of excited spin-3/2 neutrinos with $J_1$ and $J_2$
are given in Table II and Table III, respectively. Also, total decay width of excited neutrinos as a function of their mass ($m^{\star}$) is shown in Figure 1. 

\begin{table}[h]
\caption{Brancing ratios and total decay width of excited spin-3/2 neutrinos
with $J_{1}$. Here it is taken $c_{1V}=c_{1A}$=0.5 and $\Lambda=m^{\star}$.}

\begin{tabular}{|c|c|c|c|c|}
\hline 
$m^{\star}(GeV)$ & $\Gamma(GeV)$ & $\%BR($$\nu^{\star}\rightarrow\nu\gamma)$ & $\%BR($$\nu^{\star}\rightarrow\nu Z)$ & $\%BR($$\nu^{\star}\rightarrow eW)$\tabularnewline
\hline 
\hline 
300 & 1.21 & 24.0 & 34.4 & 41.6\tabularnewline
\hline 
500 & 3.89 & 12.5 & 39.0 & 48.5\tabularnewline
\hline 
750 & 11.11 & 6.5 & 41.2 & 52.3\tabularnewline
\hline 
1000 & 24.61 & 3.9 & 42.1 & 54.0\tabularnewline
\hline 
1500 & 78.89 & 1.8 & 42.8 & 55.3\tabularnewline
\hline 
2000 & 183.50 & 1.1 & 43.1 & 55.9\tabularnewline
\hline 
2500 & 355.20 & 0.7 & 43.2 & 56.1\tabularnewline
\hline 
3000 & 611.00 & 0.5 & 43.3 & 56.2\tabularnewline
\hline 
\end{tabular}
\end{table}

\begin{table}[th]
\caption{Brancing ratios and total decay width of excited spin-3/2 neutrinos
with $J_{2}$. Here it istaken $c_{2V}=c_{2A}$=0.5 and $\Lambda=m^{\star}$.}

\begin{tabular}{|c|c|c|c|c|}
\hline 
$m^{\star}(GeV)$ & $\Gamma(GeV)$ & $\%BR($$\nu^{\star}\rightarrow\nu\gamma)$ & $\%BR($$\nu^{\star}\rightarrow\nu Z)$ & $\%BR($$\nu^{\star}\rightarrow eW)$\tabularnewline
\hline 
\hline 
300 & 0.55 & 8.8 & 38.4 & 52.8\tabularnewline
\hline 
500 & 2.71 & 3.0 & 41.8 & 55.3\tabularnewline
\hline 
750 & 9.31 & 1.3 & 42.7 & 56.0\tabularnewline
\hline 
1000 & 22.21 & 0.7 & 43.0 & 56.2\tabularnewline
\hline 
1500 & 75.26 & 0.3 & 43.3 & 56.4\tabularnewline
\hline 
2000 & 178.7 & 0.2 & 43.4 & 56.5\tabularnewline
\hline 
2500 & 349.2 & 0.1 & 43.4 & 56.5\tabularnewline
\hline 
3000 & 603.6 & 0.1 & 43.4 & 56.5\tabularnewline
\hline 
\end{tabular}
\end{table}

\begin{figure}[h]
\includegraphics{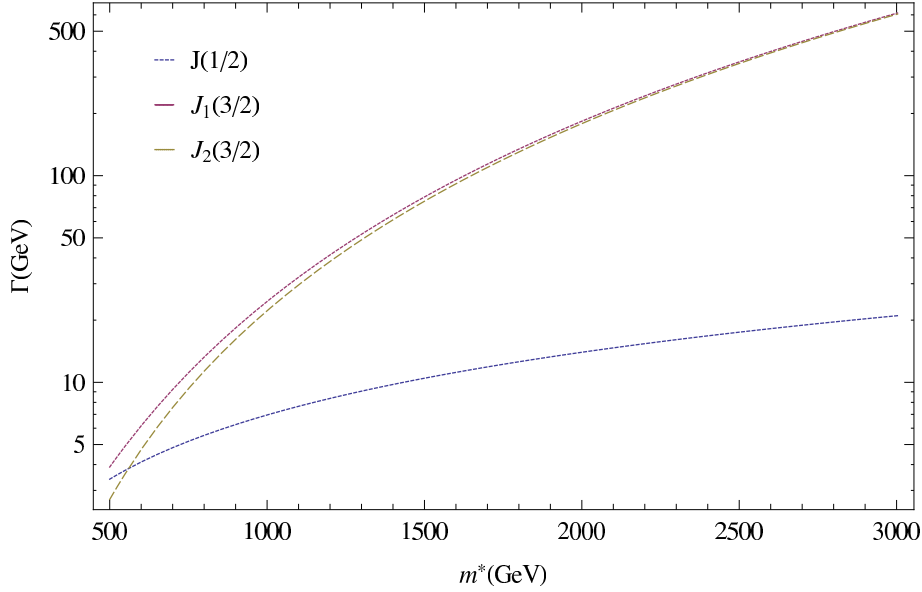}

\caption{Total decay width of excited neutrinos according to their mass. Here,
it is taken $\Lambda=m^{\star}$, $f=-f^{\prime}=1$ for excited spin-1/2
neutrinos and $c_{iV}=c_{iA}=0.5 (i=1,2)$ for excited spin-3/2 neutrinos for
the two phenomenological currents. }
\end{figure}

\section{Single production at ep collider}

The excited spin-1/2 and spin-3/2 neutrinos can be produced singly
at future $ep$ colliders via $t-$channel $W$ exchange. In our calculations
we use the program CALCHEP \cite{mycitation-27}.The Feynman diagrams
for the subprocess $e^{-}q\rightarrow\nu^{\star}q^{\prime}$ and $e^{-}\bar{q^{\prime}\rightarrow\nu^{\star}\bar{q}}$
are shown in Figure 2. 

\begin{figure}[h]
\caption{Feynman diagram }

\includegraphics[scale=1.2]{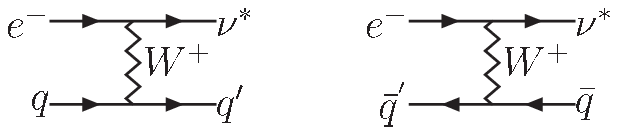}
\end{figure}

Total cross section as a function of excited neutrino mass is shown
in Figure 3 for the center of mass energies $\sqrt{s}=1.3$ TeV and
$\sqrt{s}=1.98$ TeV.

\begin{figure}[H]
\includegraphics[scale=0.7]{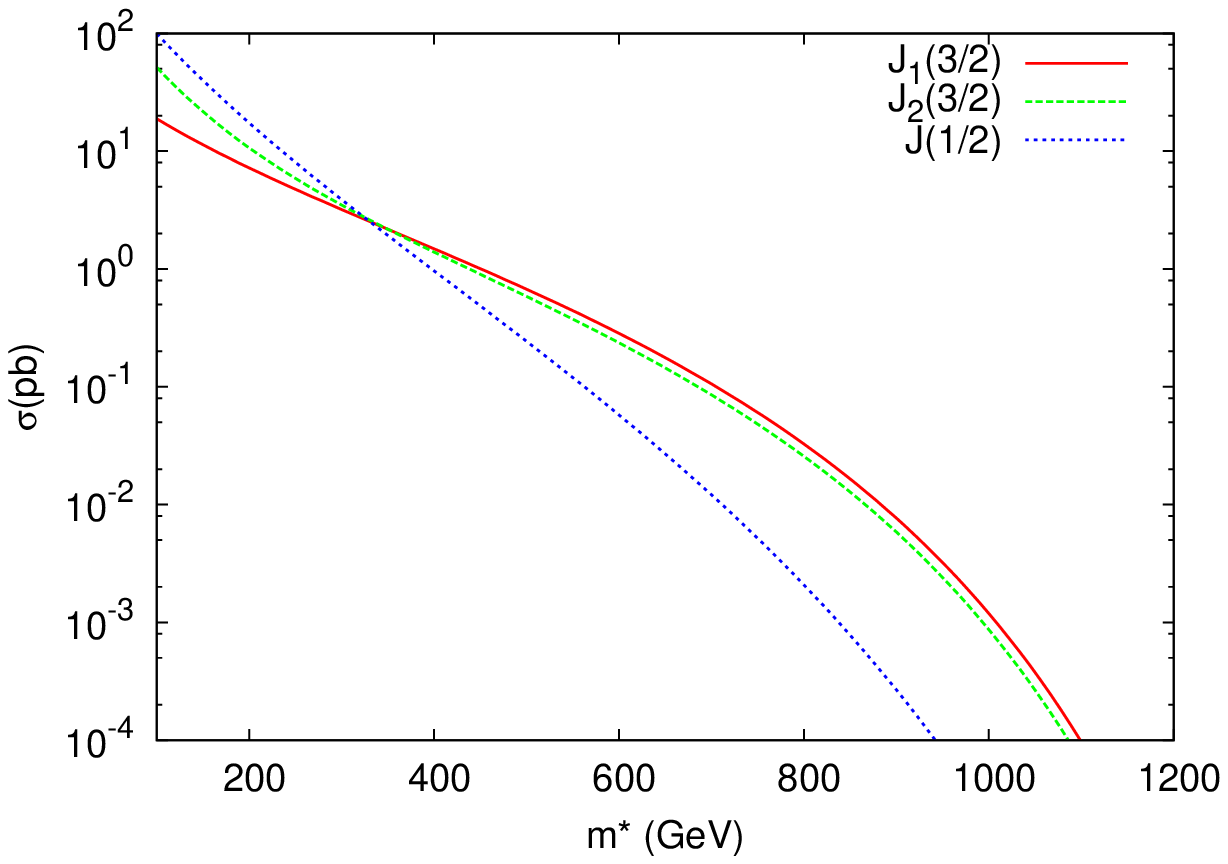}\includegraphics[scale=0.7]{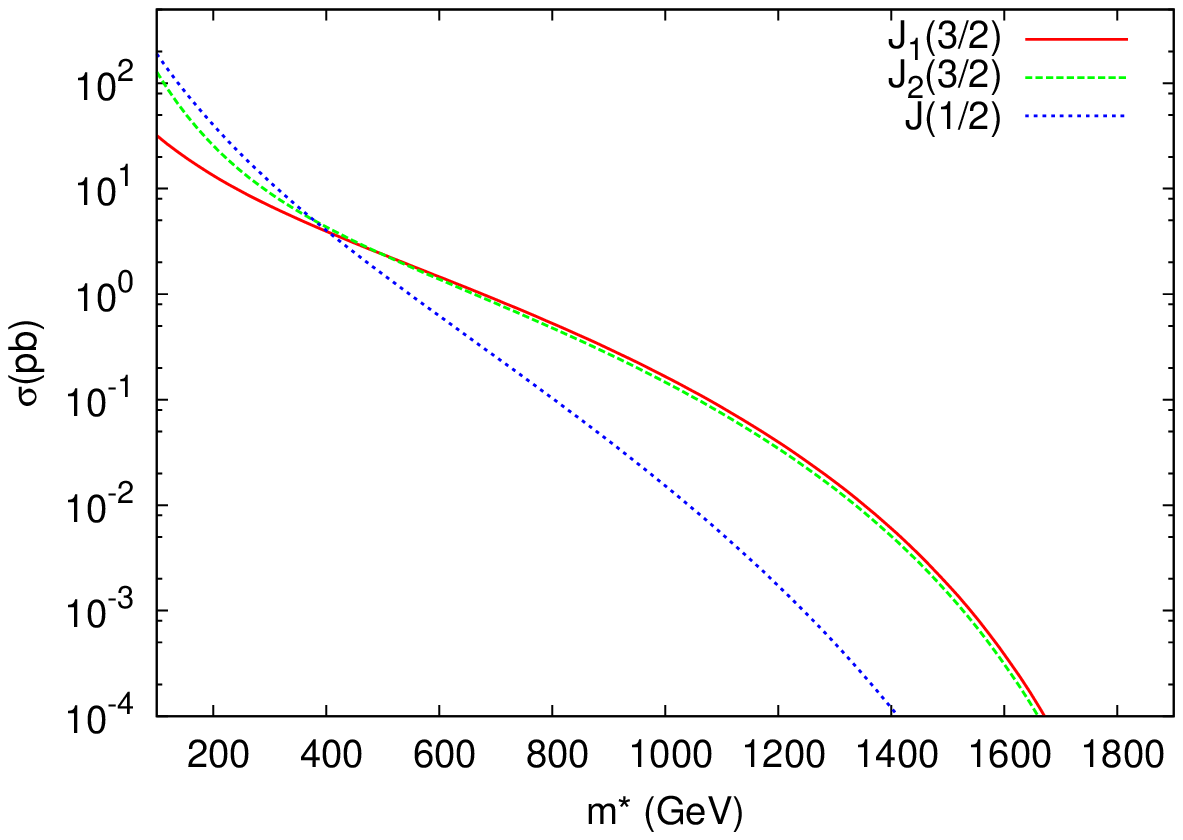}

\caption{Cross sections for the excited neutrino production with $\Lambda=m^{\star}$
and $f=-f^{\prime}$ for spin-1/2 ones and $c_{iV}=c_{iA}=0.5\:(i=1,2)$
for spin-3/2 ones at $ep$ collider at $\sqrt{s}=1.3$ TeV and $\sqrt{s}=1.98$
TeV.}
\end{figure}

In our analysis we chose the $\nu^{\star}\rightarrow eW$ mode because
of the high branching ratio of the charged current decay channel.
We consider the $ep\rightarrow\nu^{\star}X\rightarrow W^{+}e^{-}X$
process and put some kinematical cuts for the final state detectable
particles. We deal with the subprocess $e^{-}q(\bar{q})^{\prime}\rightarrow W^{+}e^{-}q^{\prime}(\bar{q})$
and impose the acceptance cuts 

\begin{equation}
p_{T}^{e,q}>20\: GeV,
\end{equation}

\begin{equation}
|\eta^{e,q}|<2.5
\end{equation}

After applying these cuts we obtained the SM background cross section
for the process $ep\rightarrow\nu^{\star}X\rightarrow e^{-}W^{+}X$
as $\sigma_{B}=0.334$ pb for $\sqrt{s}=1.3$ TeV and $\sigma_{B}=0.928$
pb for $\sqrt{s}=1.98$ TeV. In order to discriminate the excited
neutrino signal we plot the invariant mass distributions for the $eW$
system for the masses $m^{\star}=400,\,500,\,600$ GeV at $\sqrt{s}=1.3$
TeV and for the masses $m^{\star}=700,\,800,\,900$ GeV at $\sqrt{s}=1.98$
TeV in Figure 4 and Figure 5, respectively. 

\begin{figure}[H]
\includegraphics[scale=0.65]{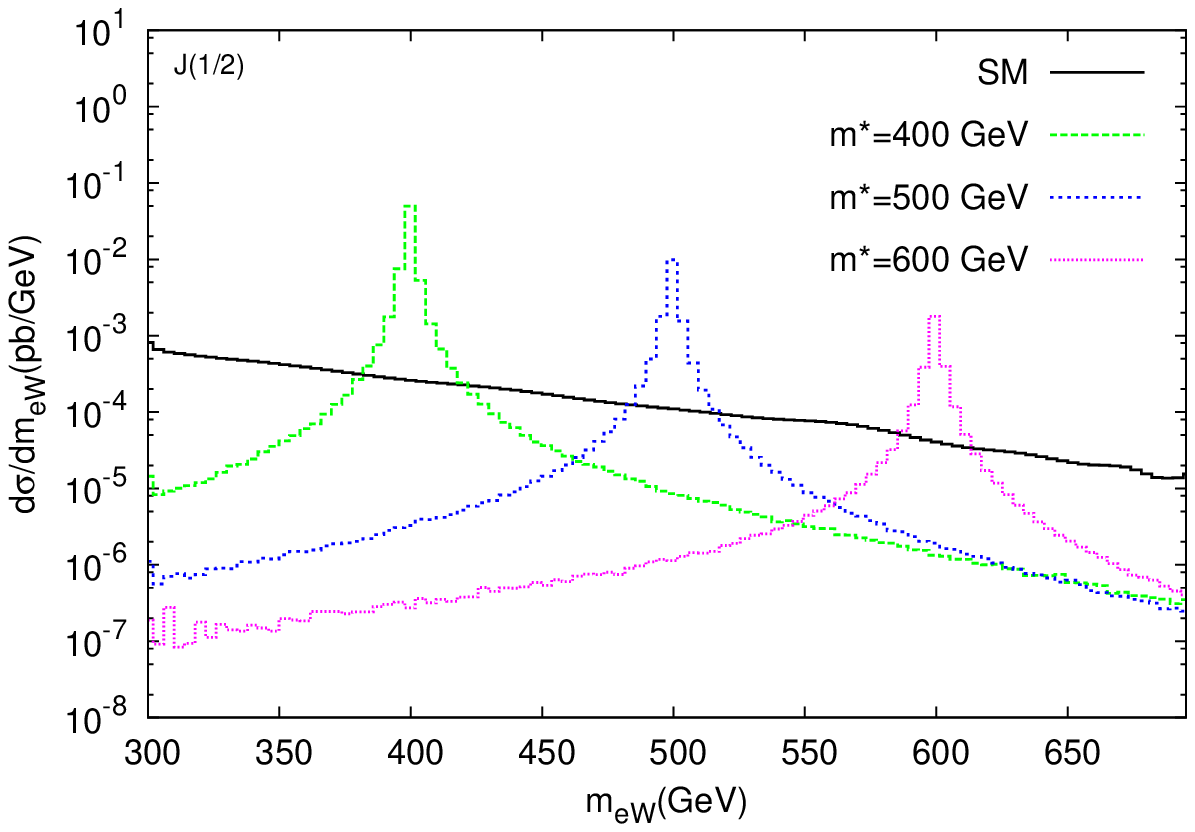}\includegraphics[scale=0.65]{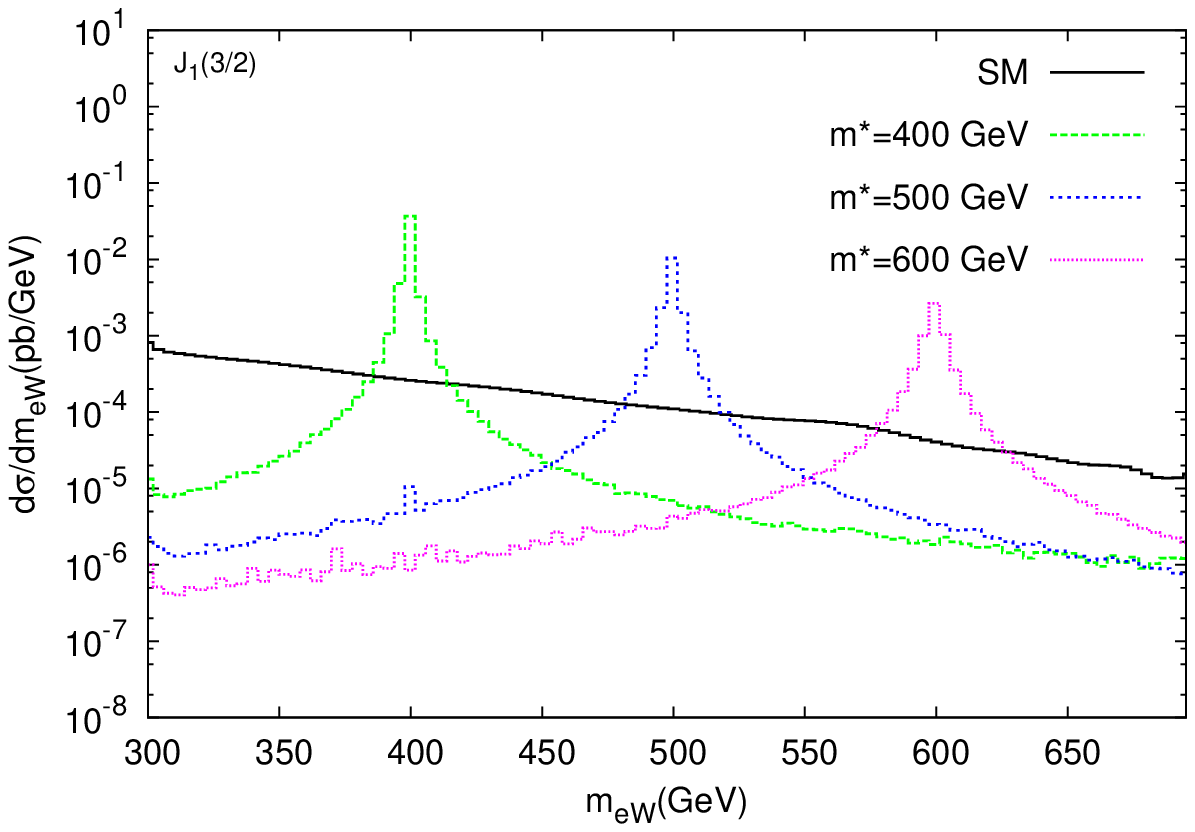}

\includegraphics[scale=0.65]{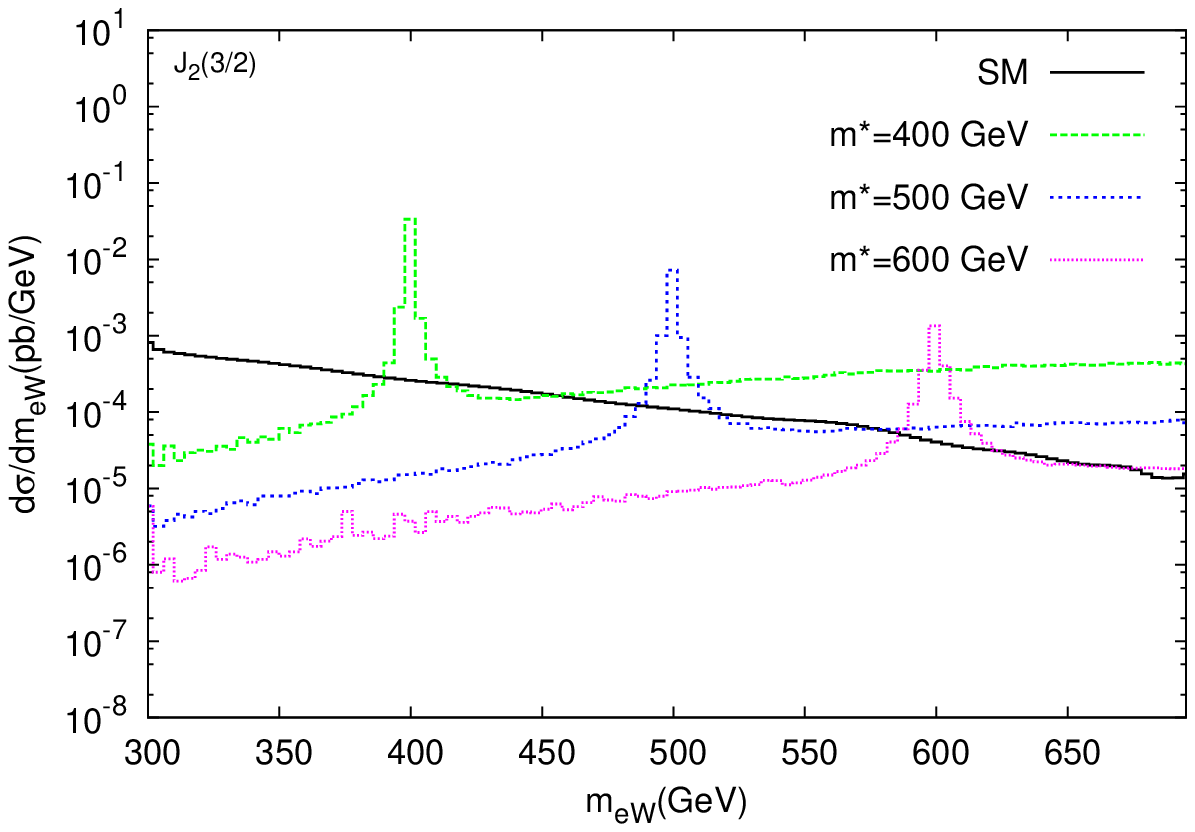}

\caption{Invariant mass distributions of $eW$ system for the single production
of excited spin-1/2 and excited spin-3/2 neutrinos with $J_{1}$ and
$J_{2}$ for $\sqrt{s}=1.3$ TeV.}
\end{figure}

\begin{figure}[H]
\includegraphics[scale=0.65]{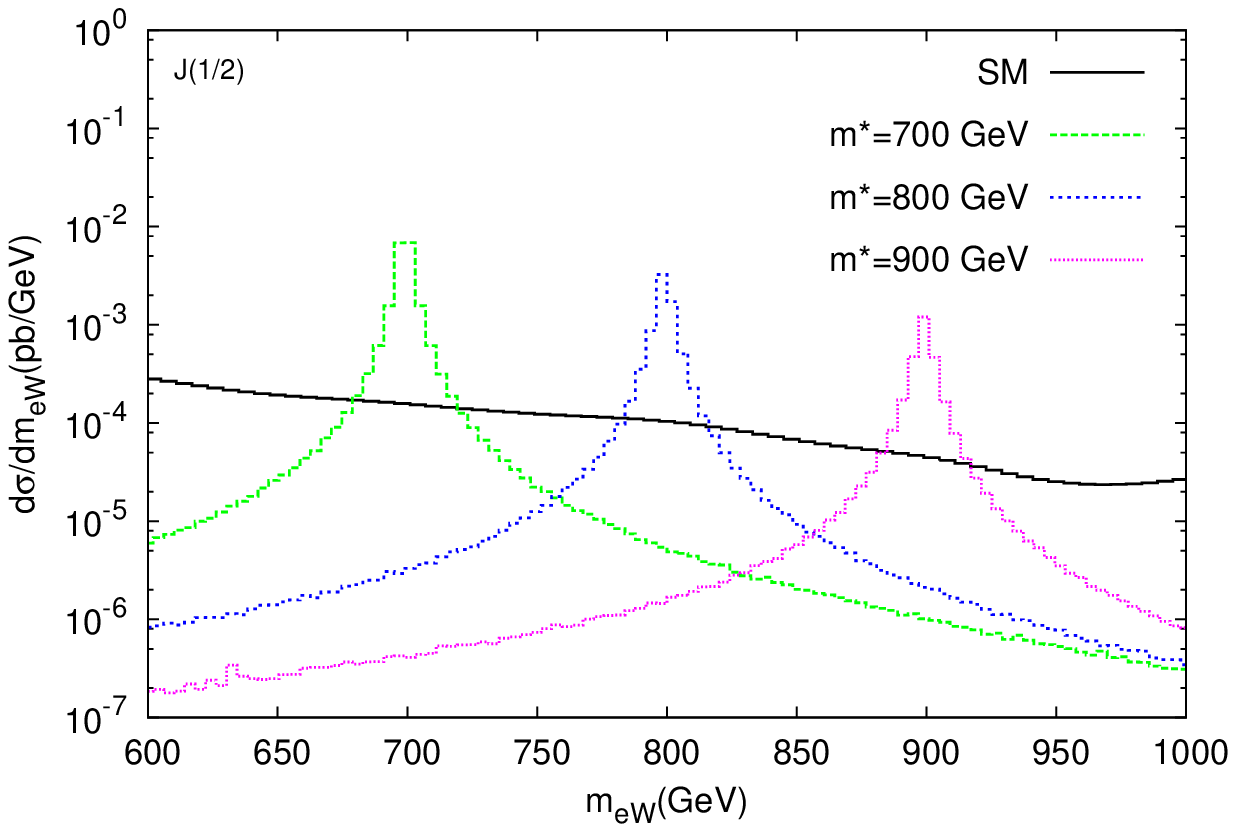}\includegraphics[scale=0.65]{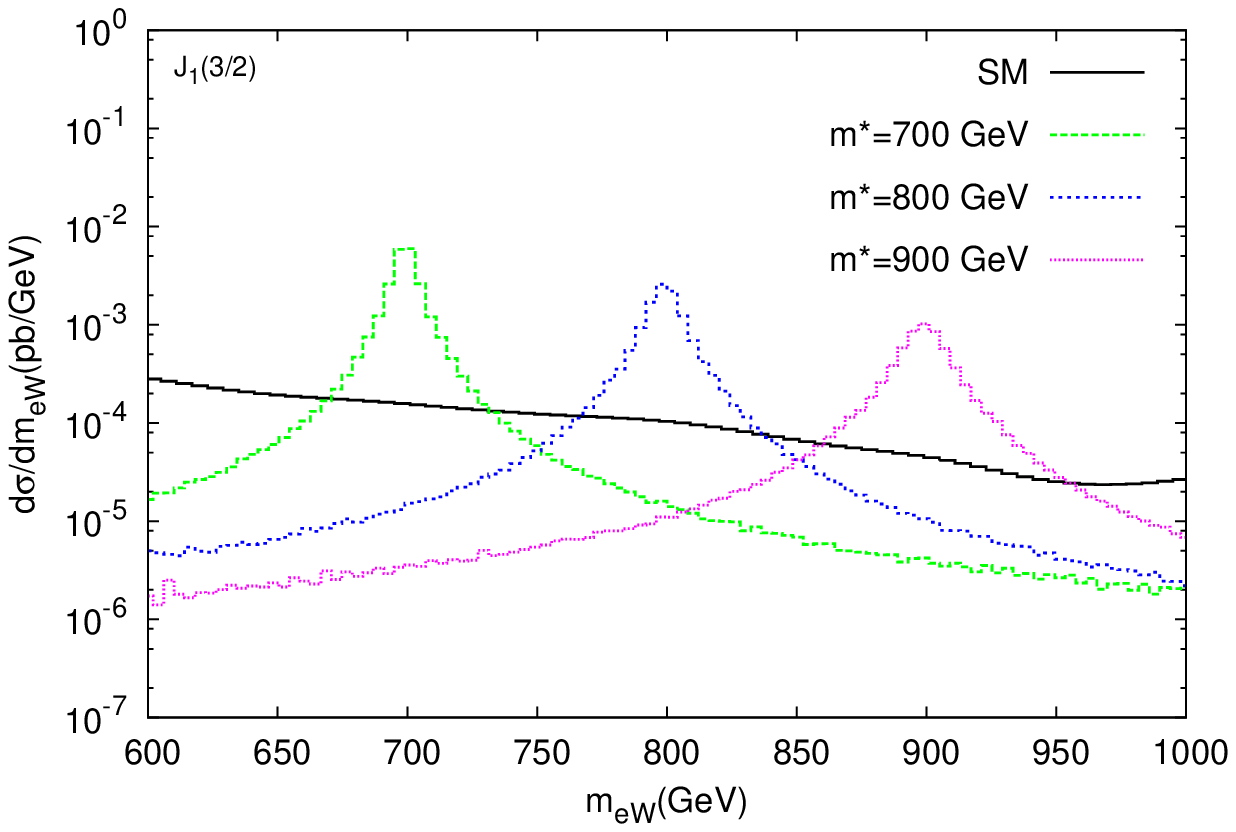}

\includegraphics[scale=0.65]{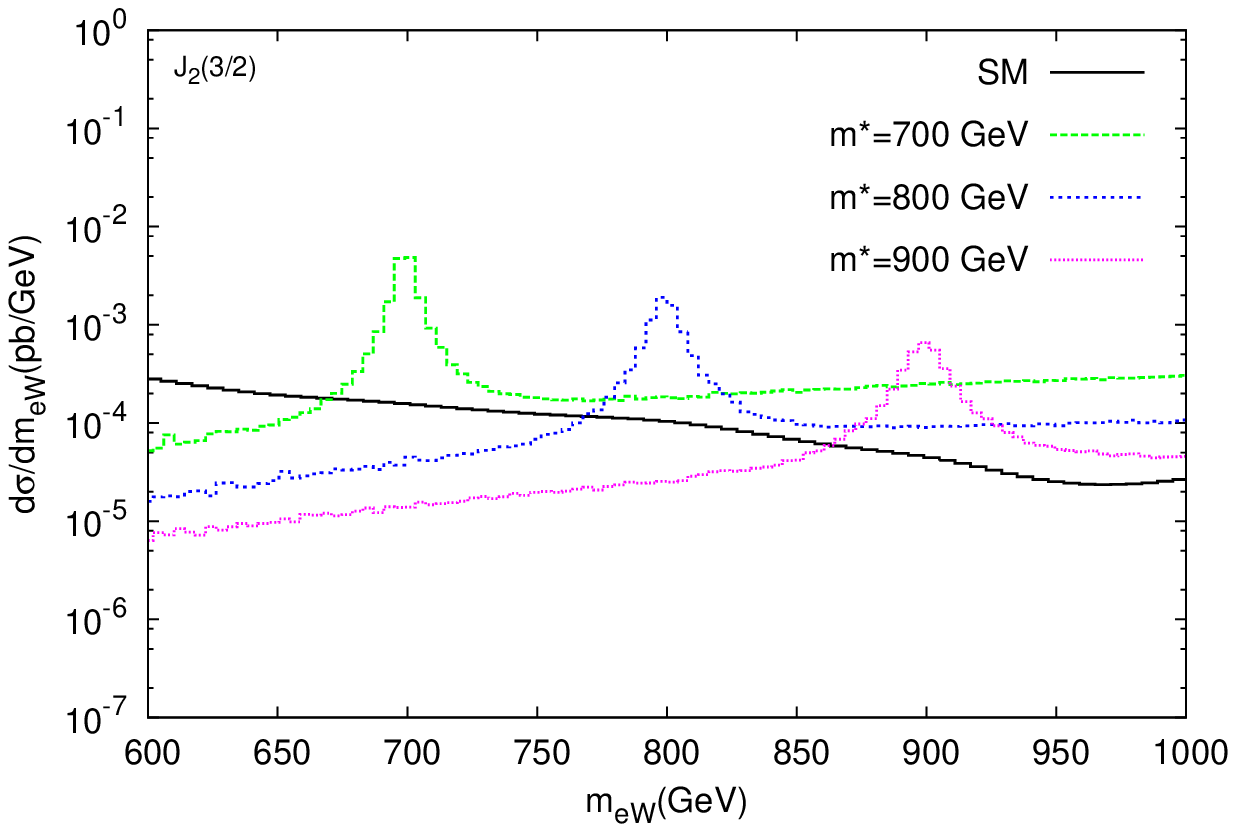}

\caption{Invariant mass distributions of $eW$ system for the single production
of excited spin-1/2 and excited spin-3/2 neutrinos with $J_{1}$ and
$J_{2}$ for $\sqrt{s}=1.98$ TeV.}
\end{figure}

We plot the rate of $\sigma_{B+S}/\sigma_{B}$ as a function of excited
neutrino mass in Figure 6 to examine the contribution of excited neutrinos
to the process $e^{-}q(\bar{q})^{\prime}\rightarrow W^{+}e^{-}q^{\prime}(\bar{q})$
and also, to investigate the separation of different excited neutrino
models. Here $\sigma_{B+S}$ corresponds the cross section calculated
for the presence of excited neutrino (signal) and Standard Model (background)
both, and $\sigma_{B}$ is the SM (background) cross section. In these
figures, the seperation spin-1/2, spin-3/2 with $J_{1}$ and spin-3/2
with $J_{2}$ excited neutrinos can be easily seen.

\begin{figure}[H]
\includegraphics[scale=0.65]{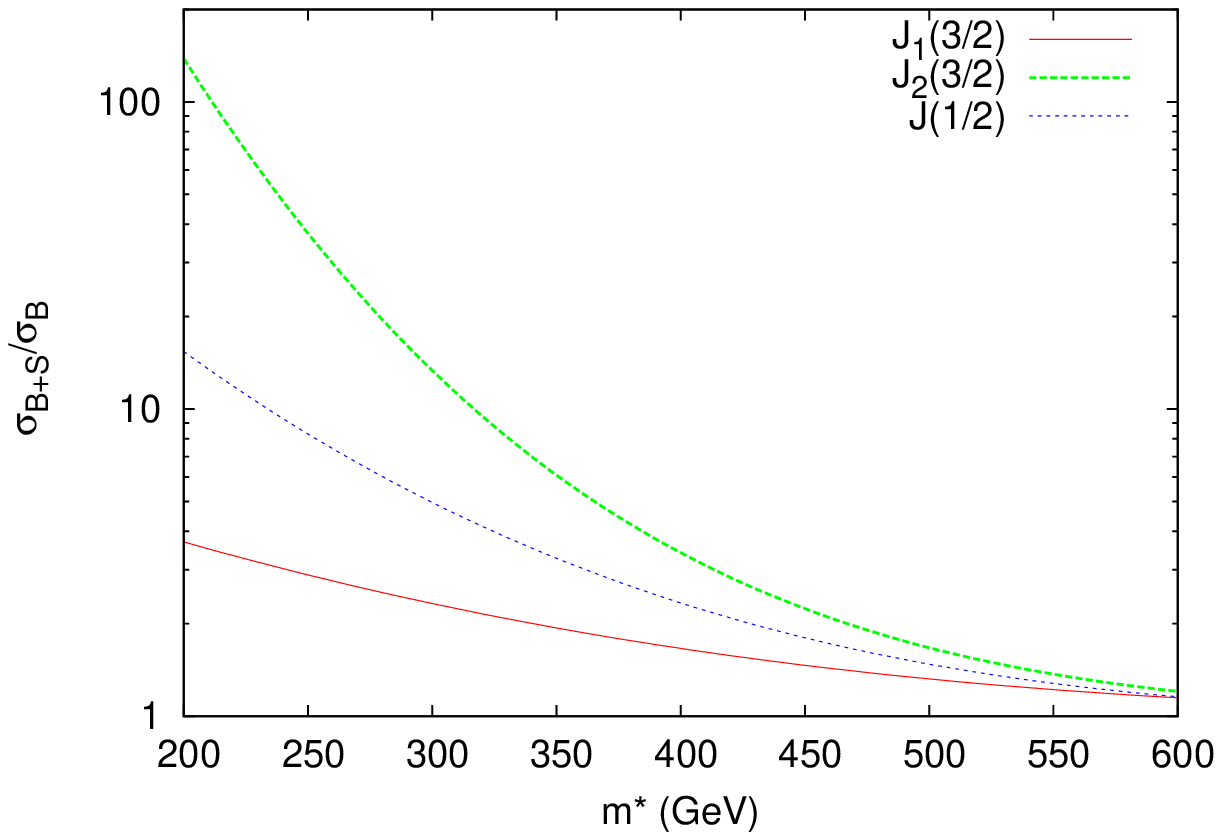}\includegraphics[scale=0.65]{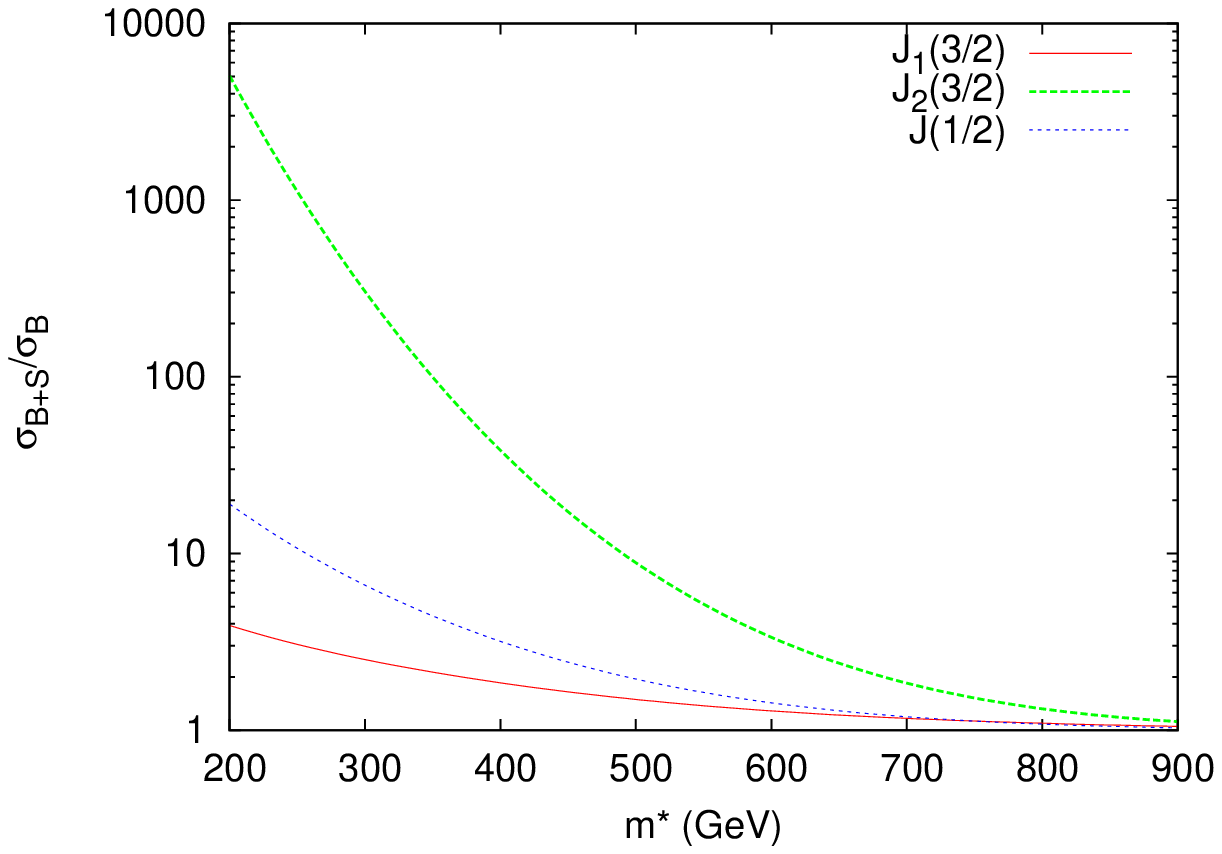}

\caption{$\sigma_{B+S}/\sigma_{B}-m^{\star}$ plots for $\sqrt{s}=1.3$ TeV
(left) and $\sqrt{s}=1.98$ TeV (right). }
\end{figure}

In order to get accessible limits for the excited neutrinos at high
energy ep collider, we plot the contour graphics for excited neutrinos
with spin-1/2 and spin-3/2. We choose the $W$ boson decay as $W\rightarrow2j$.
Here we consider the statistical significance 

\begin{equation}
SS=\frac{\sigma_{S}}{\sqrt{\sigma_{B}}}\sqrt{L_{int}}
\end{equation}

Here $L_{int}$ is the integrated luminosity of the $ep$ collider
and we choose $L_{int}=100$ $fb^{-1}$ as the LHeC design luminosity.
Our results for the $SS$ are shown in Table IV and Table V. 

\begin{table}[h]
\caption{Statistical significance $SS$ for $ep$ collider with $\sqrt{s}=1.3$
TeV for excited spin-1/2 neutrinos and excited spin-3/2 neutrinos
with $J_{1}$ and $J_{2}$. }

\begin{tabular}{|c|c|c|c|}
\hline 
$m^{\star}$(GeV) & $SS(J(1/2))$ & $SS(J_{1}(3/2))$ & $SS(J_{2}(3/2))$\tabularnewline
\hline 
\hline 
400 & 110.2 & 75.4 & 135.6\tabularnewline
\hline 
500 & 25.5 & 30.7 & 30.0\tabularnewline
\hline 
600 & 5.5 & 11.9 & 7.9\tabularnewline
\hline 
700 & 1.02 & 4.2 & 2.2\tabularnewline
\hline 
\end{tabular}
\end{table}

\begin{table}[h]
\caption{Statistical significance $SS$ for $ep$ collider with $\sqrt{s}=1.98$
TeV for excited spin-1/2 neutrinos and excited spin-3/2 neutrinos
with $J_{1}$ and $J_{2}$. }

\begin{tabular}{|c|c|c|c|}
\hline 
$m^{\star}$(GeV) & $SS(J(1/2))$ & $SS(J_{1}(3/2))$ & $SS(J_{2}(3/2))$\tabularnewline
\hline 
\hline 
600 & 56.3 & 51.0 & 235.9\tabularnewline
\hline 
700 & 22.4 & 28.0 & 76.5\tabularnewline
\hline 
800 & 8.8 & 15.1 & 28.9\tabularnewline
\hline 
900 & 3.3 & 8.04 & 12.0\tabularnewline
\hline 
1000 & 1.2 & 4.2 & 5.3\tabularnewline
\hline 
\end{tabular}
\end{table}

Concerning the criteria $SS\geqslant3$ we plot the $c_{iv}-c_{iA}$
(i=1,2) contour plot for excited spin-3/2 neutrinos for the two phenomenological
currents and, $f-f^{\prime}$ contour plot for the excited spin-1/2
neutrinos. In Figure 7 and Figure 8, we choose the excited neutrino
mass $m^{\star}=400$ GeV for the analysis at $\sqrt{s}=1.3$ TeV
and $m^{\star}=800$ GeV for the analysis at $\sqrt{s}=1.98$ TeV.
We see from these figures the allowed regions for the $c_{iv}-c_{iA} (i=1,2)$
and $f-f^{\prime}$ couplings for the masses $m^{\star}=400$
GeV at $\sqrt{s}=1.3$ TeV and $m^{\star}=800$ GeV at $\sqrt{s}=1.98$
TeV. The values which we chose in our calculations for the coupling
parameters ($c_{iV}=c_{iA}=0.5$ for the excited spin-3/2 neutrinos
and $f=-f^{\prime}=1$ for the excited spin-1/2 neutrinos) are compatible
with the contour plots.

\begin{figure}
\includegraphics[scale=0.65]{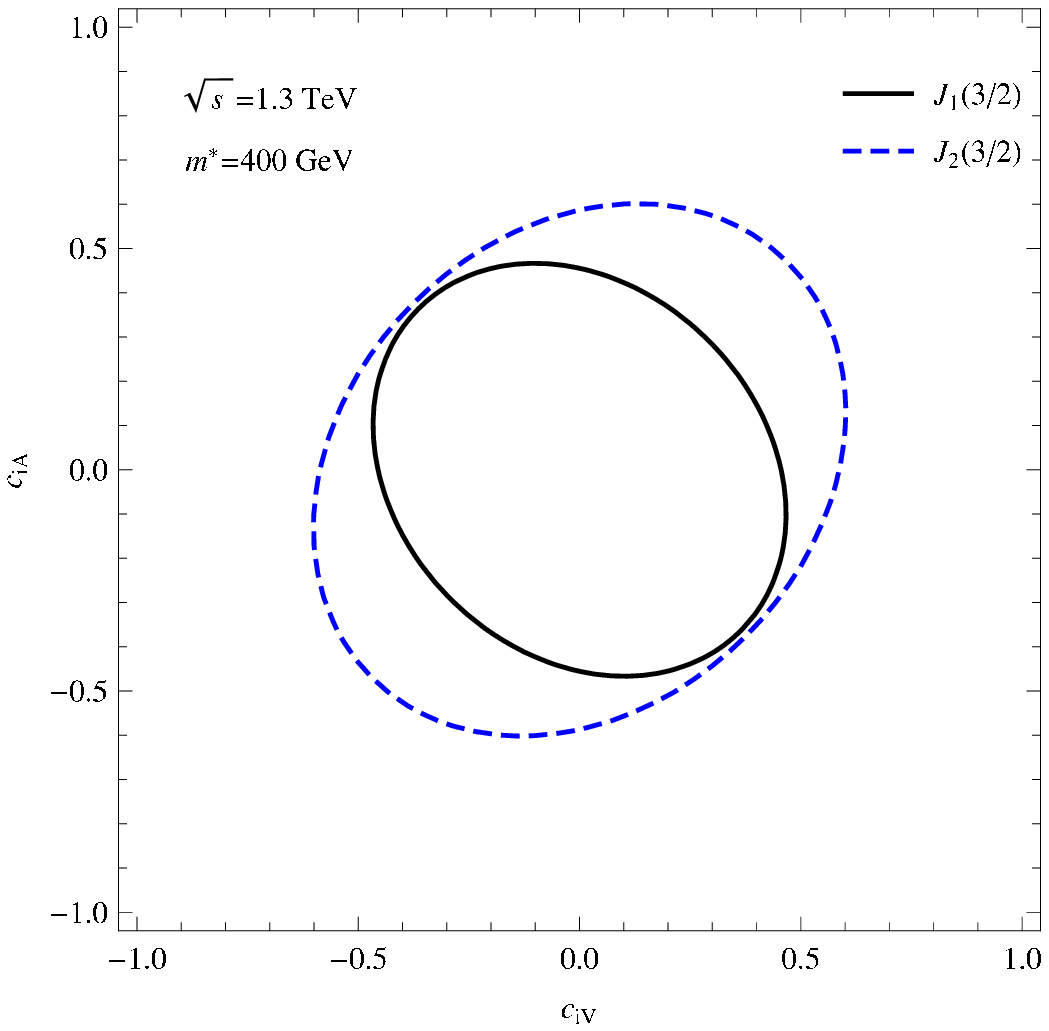}\includegraphics[scale=0.65]{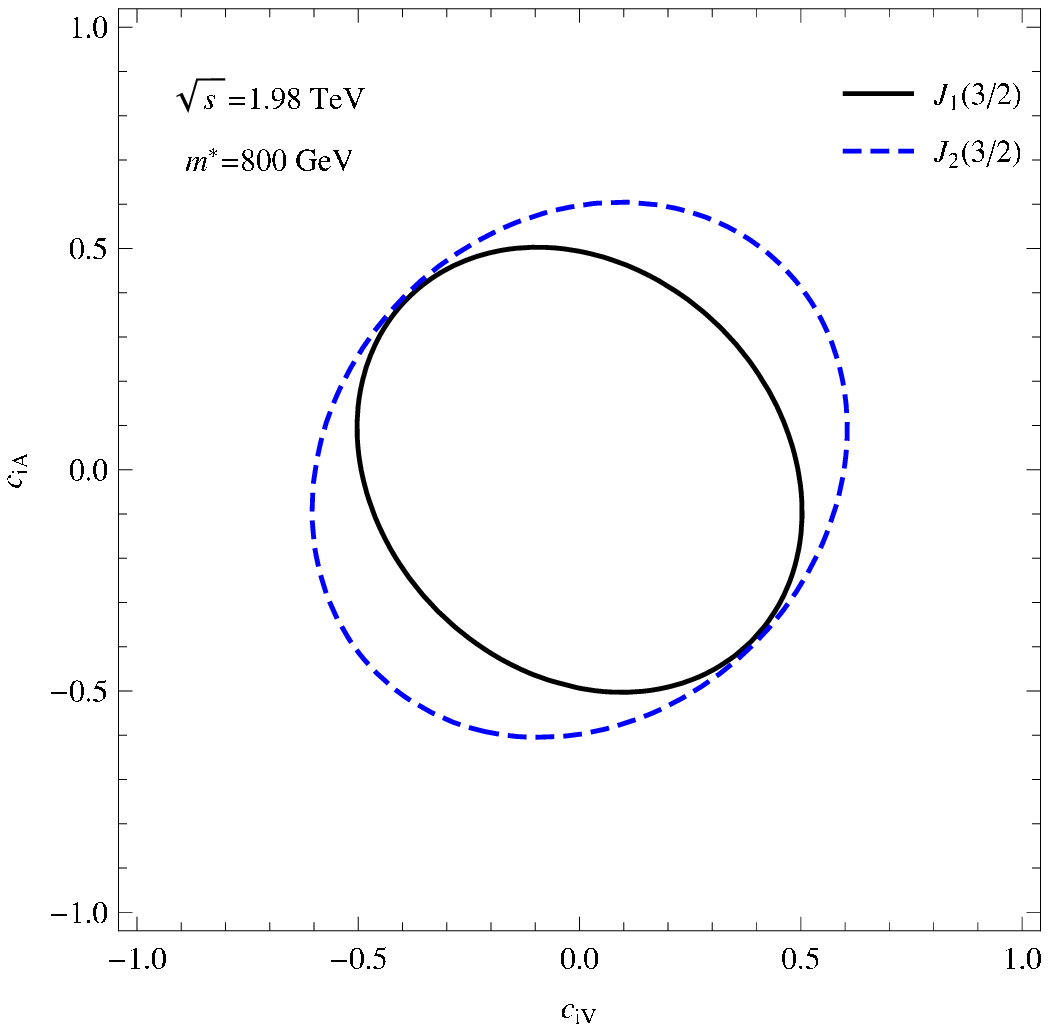}

\caption{Contour plots for excited spin-3/2 neutrinos for the $J_{1}$and $J_{2}$.}
\end{figure}

\begin{figure}
\includegraphics[scale=0.65]{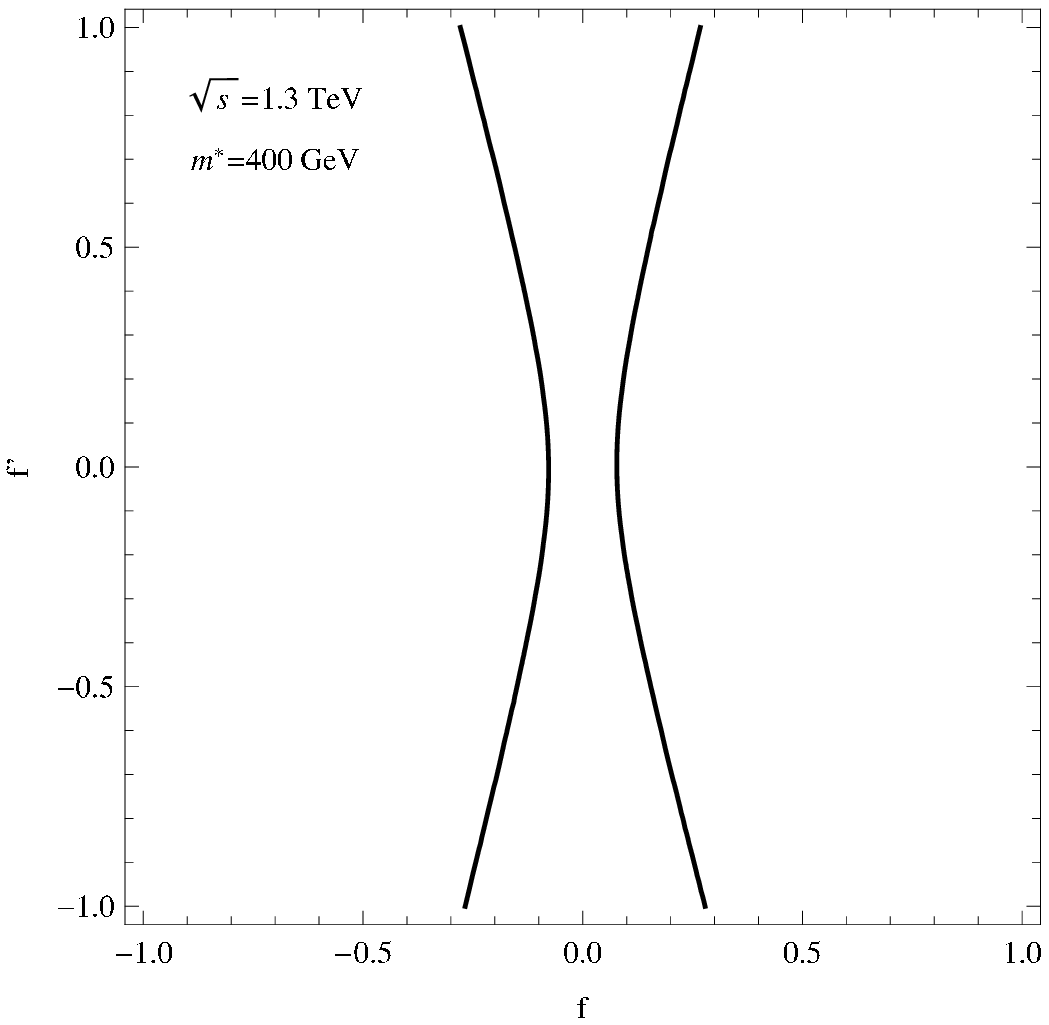}\includegraphics[scale=0.65]{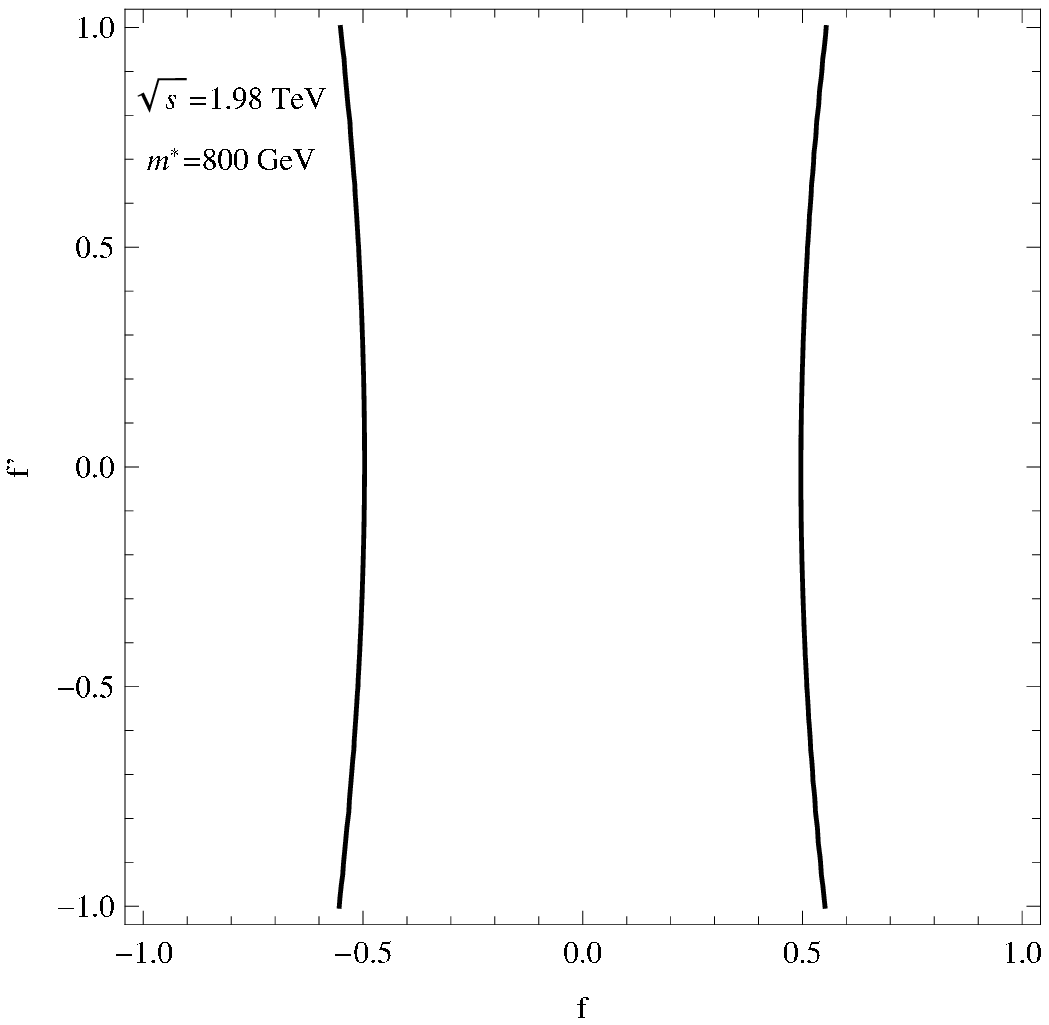}

\caption{Contour plots for excited spin-1/2 neutrinos.}
\end{figure}

\section{Conclusion}

We searched for the excited spin-3/2 neutrino signal at lepton-hadron
collider LHeC for two different center of mass energies. We used two
different phenomenological current for the spin-3/2 excited neutrinos,
and we use the same value of $c_{iV},c_{iA}(i=1,2)$ couplings. Since
there is no theoretical prediction for the single production of excited
neutrinos and the effective currents have unknown couplings, we didn't
consider the interference between the currents.

In a more detailed calculation one can find an important parameter
space in which the interference terms could be important. We also
deal with the spin-1/2 excited neutrinos for comparision. Our analysis
show that the spin-1/2 and spin3/2 excited neutrino signals discrimination
is apparent at next $ep$ colliders. Here we only take into account
the effective currents describing the gauge interactions of excited
and standard particles. It is possible to include the contact interactions
which may enlarge the mass and coupling limits. 

Excited neutrinos with different spins would manifest themselves in
three famlies. Here, we only investigate for the excited electron
neutrino. It is also possible to make the same analysis for excited
muon neutrinos. Single production of excited muon neutrinos is possible
at muon-hadron colliders. Physics of $\mu p$ colliders was studied
in \cite{key-1}. One can find the main parameters of $FCC$-based
$\mu p$ collider in\cite{key-2}.

\end{document}